\documentclass[aps,prl,reprint,superscriptaddress]{revtex4-1}

\usepackage{longtable,gensymb}
\usepackage{morefloats}
\usepackage{color}
\usepackage{graphicx,amsfonts,amsbsy}
\usepackage{amsmath,amsfonts,amsthm,amssymb}
\usepackage{appendix}
\usepackage{mathtools}
\usepackage{makeidx}
\usepackage{url}
\usepackage{verbatim}
\usepackage[bookmarksnumbered,pdfpagelabels=true,plainpages=false,colorlinks=true,linkcolor=blue,citecolor=blue,urlcolor=blue]{hyperref}
\usepackage[rightcaption]{sidecap}
\usepackage{array}
\usepackage{booktabs}
\usepackage{multirow}
\usepackage{tabularx}
\usepackage{cancel,soul,ulem}

\usepackage{mathrsfs}

\newcommand{\bs}[1]{\boldsymbol{#1}}

\begin{document}
\title{Magnetostatic interaction between Bloch Point nanospheres}
\author{Cristobal Zambrano-Rabanal}
\affiliation{Departamento de Ciencias F\'isicas, Universidad de La Frontera, Casilla 54-D, Temuco, Chile}
\author{Boris Valderrama}
\affiliation{Facultad de Física, Pontificia Universidad Católica de Chile, Casilla 306, Santiago, Chile}
\author{Felipe Tejo}
\affiliation{Escuela de Ingenier\'ia, Universidad Central de Chile, Avda. Santa Isabel 1186, 8330601 Santiago, Chile}
\author{Ricardo Gabriel Elías}
\affiliation{Departamento de Física, Universidad de Santiago de Chile (USACH), Avda. Víctor Jara 3493, Santiago, Chile}
\author{Alvaro S. Nunez}
\affiliation{Departamento de Física, FCFM, Universidad de Chile, Santiago, Chile.}
\affiliation{Centro  de nanociencia y nanotecnología CEDENNA, Avda. Ecuador 3493, Santiago, Chile}
\author{Vagson L. Carvalho-Santos}
\affiliation{Departamento de F\'isica, Universidade Federal de Vi\c cosa, 36570-900, Vi\c cosa, Brazil}
\author{Nicolás Vidal-Silva}
\email{nicolas.vidal@ufrontera.cl}
\affiliation{Departamento de Ciencias F\'isicas, Universidad de La Frontera, Casilla 54-D, Temuco, Chile}

\begin{abstract}
Three-dimensional topological textures have become a topic of intense interest in recent years. Through analytical calculations, this work determines the magnetostatic field produced by a Bloch Point (BP) singularity confined in a magnetic nanosphere. It is observed that BPs hosted in a nanosphere generate magnetic fields with quadrupolar nature. This finding is interesting because it shows the possibility of obtaining quadrupole magnetic fields with just one magnetic particle, unlike other propositions considering arrays of magnetic elements to generate this kind of field. The obtained magnetostatic field allows us to determine the interaction between two BPs as a function of the relative orientation of their polarities and the distance between them. It is shown that depending on the rotation of one BP related to the other, the magnetostatic interaction varies in strength and character, being attractive or repulsive. The obtained results reveal that the BP interaction has a complex behavior beyond topological charge-mediated interaction.  
\end{abstract}

\maketitle
 
\begin{section}{Introduction}

The analysis of the properties of solitons is a cornerstone in current research regarding technological applications based on the control of magnetic quasiparticles \cite{Parkin-Nat,Fert-Nat,Hrcak,Grolier,Marrows-APL}. The main feature that allows the use of solitonic magnetic patterns as information carriers resides in their strong stability ensured by a topological protection \cite{Rajaraman}. Therefore, the focus of applied research is on describing the fundamental properties of topological textures and their current-driven motion. Up to recent years, the interest was centered in studying solitons lying in quasi-2D-systems, such as skyrmions \cite{Sampaio-Nat}, skyrmioniums \cite{Skyrmionium}, biskyrmions \cite{biskyrmion}, and bimerons \cite{bimerons-1,bimerons-2,bimerons-3,bimerons-4}. Nevertheless, recent advances in producing and characterizing nano and microstructures with a plethora of shapes and sizes \cite{Donnely-PRL,Amalio-SciRep,May-NatCom,Phatak-PRL} renewed the interest in describing the properties of three-dimensional (3D) magnetic profiles \cite{Oksana,Birsh-ACS,Seki-NatMat,Donnely-NatPhys,bobbers-1,bobbers-2,Tai-PRL,Liu-PRB,Castillo-PRB,Sutcliffe-PRL}. Among the 3D magnetic quasiparticles with topological protection, we can highlight the Bloch point (BP) \cite{Col-PRB,Im-NatCom}, which is a structure that presents a singular point at its center, where ferromagnetic order is destroyed \cite{Feldkeler,Doring,Galkina}. The defining property of a BP is that in a closed surface around its center, the direction of the magnetization field covers the whole solid angle an integer number of times. Theoretical and experimental works showed that these magnetic singularities appear in magnetic nanodots with perpendicular magnetic anisotopy \cite{Tai-PRL}, in ferromagnets during the process of vortex core reversal \cite{Thiaville-PRB,Hertel-PB,Hertel-PRL}, the reversal of skyrmions in confined helimagnetic structures \cite{Milde-SC}, in a bilayer of nanodots of FeGe with different chiralities \cite{Beg-SciRep}, in modulated nanowires with intrinsic Dzialoshinskii-Moriya interaction \cite{Saez-RP}, and in cylindrical magnetic nanowires as the center of a vortex domain wall \cite{Wieser-PRB,Jamet-Book,Moreno-JMMM}. 

To properly use BPs in technological applications based on their stabilization and motion, it is crucial to analyze the fundamental properties of these structures. Therefore, several interesting phenomena regarding the BPs behavior have been reported. For instance, the analysis of the magnetization resonant modes in modulated nanowires evidenced that the magnetic response to an external magnetic field can establish a strategy for detecting BPs in such systems \cite{Saez-2}. Also, from analyzing the scattering of magnons by BPs, El\'ias \textit{et al.} \cite{Elias-PRB} showed that this system is described by the same solutions of the system electron$\times$magnetic monopole. Therefore, the non-trivial topological structure of the Bloch point manifests in the propagation of spin waves, endowing them with a gauge potential that yields the emergence of spin wave vortices \cite{Carvalho-AOP,Jia-Nat} as a result of such a scattering. Another exciting property regarding the analogy between BPs and phenomena belonging to other physical contexts is the spontaneous emission of spin waves when a BP domain wall displaces in a cylindrical nanowire. In this case, when the DW velocity is above a threshold value, one can observe a Cherenkov-type DW breakdown phenomenon originating from an interaction between the spontaneously emitting SW and the BP domain wall \cite{Ma-APL}.

Because the structure of a BP depends on the boundary conditions, an interesting problem, from the fundamental point of view, is the analysis of the BP properties as a function of the geometrical parameters of a magnetic system where it is confined. Therefore, due to the BP symmetry, the sphere is a natural geometry where these topological structures can appear as a metastable static configuration. Although a formal solution describing the radial BP represents the maximum energy state, hedgehog BPs appear as intermediate metastable states in spherical nanoparticles before the appearance of the more stable twisted BPs \cite{Elias-EPL}. In this regard, if the complete magnetostatic interaction is considered for determining the BP profile, the twist angle of a BP hosted in a sphere is approximately $105\degree$  \cite{Pyly-PRB}. This swirling effect originated from the magnetostatic energy leads to a non-zero 3D topological charge, which does not occur for radial BPs \cite{Tejo-SciRep}.

Based on the above described, this work analyzes the fundamental properties of a magnetic sphere hosting a BP magnetization pattern. Through analytical calculations, we determine the magnetic field generated outside the sphere as a function of the BP helicity. We show that the magnetic field generated by a BP hosted in the sphere resembles that of a quadrupole. The strength of the quadrupolar field is modulated by a factor that depends on the material magnetic parameters, the polarity, and the helicity of the BP. These results allow us to analyze the interaction between two nanospheres having a BP as the magnetic state. In this case, we obtain the BP interaction as a function of the relative orientation between their polarities and the BP-to-BP distance. We conclude that the strength of the BP interaction depends on the rotation angle of one BP related to the other. Additionally, the relative orientation between the BP polarities also determines the character of the magnetostatic interaction, which can be attractive or repulsive. These obtained results reveal the complex behavior of the BP interaction, whose discussions should be beyond topological charge-mediated interaction.

\end{section}

\begin{section}{Magnetic field of a BP}

This section presents the analysis of the magnetostatic field generated by nanospheres whose dimensions allow the nucleation and stabilization of a BP singularity as a metastable configuration.
The performed calculations are obtained in the framework of the micromagnetism approach, where the magnetization, $\mathbf{M}$, can be considered as a continuous function depending on the position inside the magnetic body. The considered system consists of a magnetic sphere with radius $R$ made of a material having saturation magnetization $M_s$. The normalized magnetization field can be written in spherical polar coordinates as ${\bs M}(\bs{r})/M_s \equiv {\bs m}=\left(\sin\Theta({\bs r})\cos\Phi({\bs r}),\sin\Theta({\bs r})\sin\Phi({\bs r}),\cos\Theta({\bs r})\right)$. Under this framework, the magnetic profile of a BP configuration can be modeled with the ansatz \cite{Elias-EPL}

\begin{align}\label{ansatz}
    \Theta(\theta) = p\theta + \pi(1-p)/2\\
    \Phi(\phi) = q\phi+\gamma,
\end{align}

\noindent
where $\theta$ and $\phi$ are the standard polar and azimuthal angles describing the spherical coordinates, $p=\pm 1$ stands for the BP polarization, $q \in \mathbb{Z}$ corresponds to the winding number, and $\gamma$ is the helicity. Although $\gamma$ has slight variations along the radial position inside the sphere \cite{Tejo-SciRep}, we consider it a fixed value in the sphere volume. Thus, in this parametrization, the (2D) topological charge is $Q = pq$. It is worth noticing that this work focuses on determining the properties of topological magnetic textures with $q=1$ (BP). The properties of their $q=-1$ counterparts (Anti BP) will be considered in future works To better describe the magnetic textures determined from Eqs. \eqref{ansatz}, we present some types of BPs with different helicities and polarities in Fig. (\ref{fig_BP}). One can notice that while $p$ determines the direction in which the magnetic moments in the poles of the sphere point, $\gamma$ determines the direction of the magnetic vector field in the sphere equator. The analysis of Fig. \ref{fig_BP} evidences that, as expected, the magnetization vector field of a BP obeys the hairy ball theorem \cite{Hairy}, which states that any continuous tangent vector field on the sphere must have a vanishing point. 

\begin{figure}[h]

\centering
\includegraphics[width=9cm]{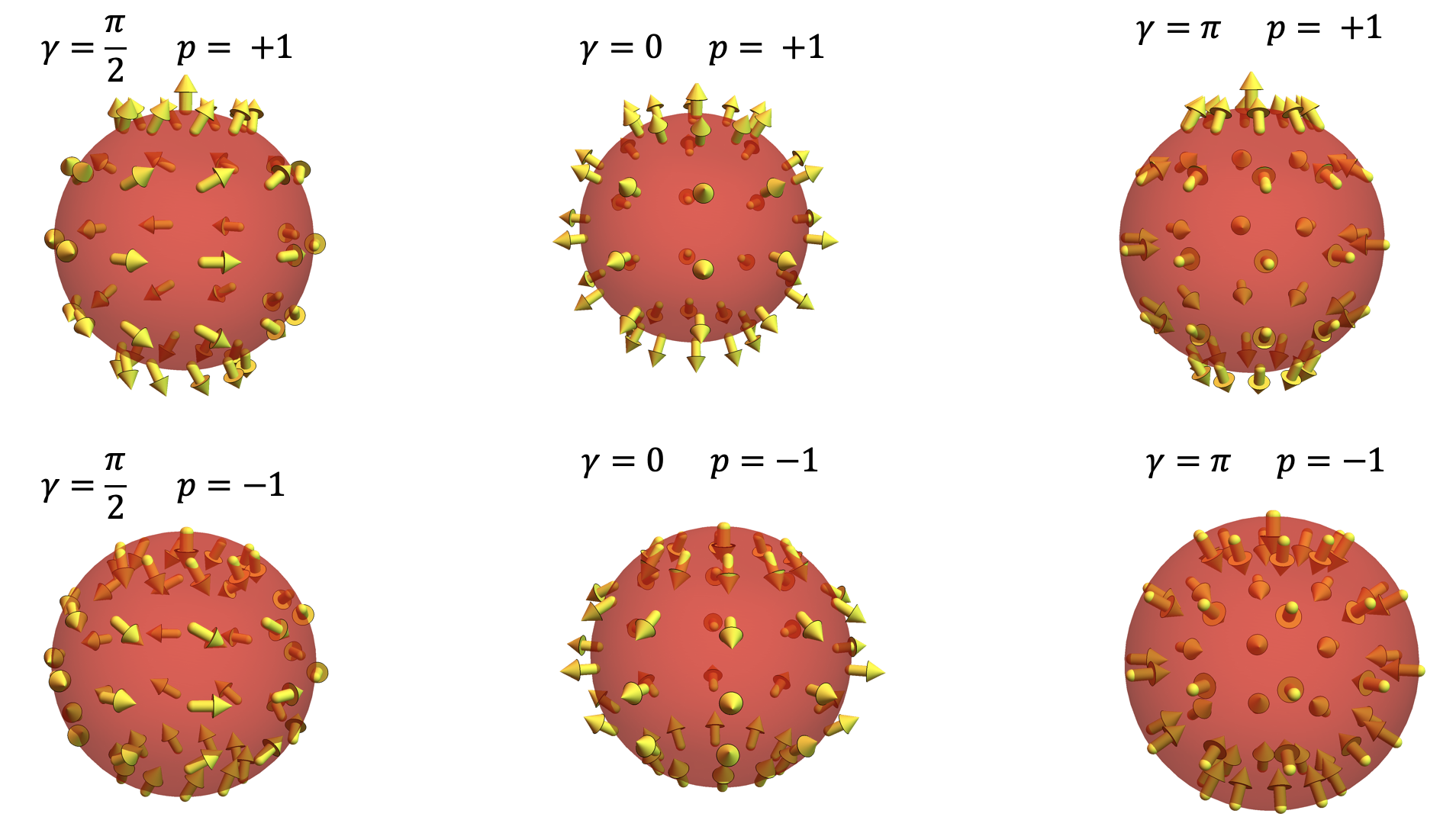}
\caption{Magnetization vector field of BP singularities confined in a nanosphere for different values of $p$ and $\gamma$.}
\label{fig_BP}
\end{figure}

One of the main objectives of the present work is determining the magnetostatic field $\bs{H}_d$, generated by a BP nucleated in a ferromagnetic nanosphere in the whole space. Indeed, although the calculation of $\bs{H}_d^{\text{in}}$ inside the sphere has already been performed \cite{Elias-EPL,Pyly-PRB}, the magnetic field, $\bs{H}_d^{\text{out}}$, generated by the BP in the region outside the sphere has not been addressed in detail. Determining the specific profile of $\bs{H}_d^{\text{out}}$ is important to better understand the fundamental physics behind BP structures and to analyze the magnetostatic interaction between two nanospheres nucleating BPs as metastable states. 

In the absence of current densities, the micromagnetism approach allows us to obtain the magnetic field as $\bs {H}_d=-\nabla U_d$, where $U_d$ is the magnetostatic potential, whose formal solution is \cite{Aharoni-Book}

\begin{equation} 
\label{Upot}
U_d(\bs r) = -\frac{1}{4\pi} \int_{V'} \frac{\nabla\cdot \bs M(\bs r')}{|\bs r-\bs r'|}dV'+\frac{1}{4\pi} \int_{ S'} \frac{\bs M(\bs r')\cdot \bs n'}{|\bs r-\bs r'|}dS', 
\end{equation}

\noindent
where $\bs n$ is the unitary vector normal to the surface, the first and second integrals in the above equation are evaluated along the sphere volume ($V$) and external surface area ($S$), respectively. The magnetostatic potential given in Eq. \eqref{Upot} can be fully calculated by expanding the Green's function $G(\bs{r},\bs{r}') = \vert\bs{r}-\bs{r}'\vert^{-1}$ into the spherical harmonic basis (see details in the Appendix \ref{SupMat}). Therefore, after some algebra, we obtain the magnetostatic potential generated by an isolated BP confined in a nanometric sphere, given by

\begin{align}
\label{Udin}
\nonumber U_d^{\text{in}}(\bs r)= \frac{M_s}{24}\Big(9pr -8pR+15r\cos\gamma\\
-16R\cos\gamma+3r\cos 2\theta(p-\cos\gamma)\Big)\,,
\end{align}

\noindent and

\begin{align}
\label{Udout}U_d^{\text{out}}(\bs r)= M_s\frac{R^4}{12 r^3}(p-\cos\gamma)(3\cos^2\theta-1)\,,
\end{align}

\noindent
where the superscripts \textit{in} and \textit{out} are associated with the magnetostatic potential inside and outside the sphere, respectively. In this context, from the definition of the demagnetizing field, we have that

\begin{align}\label{Hdall}
   \bs H_d(\bs r)= \begin{dcases}
        \bs H_d^{\text{in}}(\bs r)=-\nabla U_d^{\text{in}}(\bs r) & \text{for } r\leq R \\
       \bs H_d^{\text{out}}(\bs r)=-\nabla U_d^{\text{out}}(\bs r) & \text{for } r> R  \,.
    \end{dcases}
\end{align}

\noindent
The substitution of Eqs. \eqref{Udin} and \eqref{Udout} in Eq. \eqref{Hdall} yields the magnetostatic field inside and outside of a sphere hosting a BP, respectively given by

\begin{align}
\label{Hmagin}
\nonumber\bs H_d^{\text{in}}(\bs r)=-\frac{M_s}{8}\Big[\left(3p+5\cos\gamma+(p-\cos\gamma)\cos2\theta\right)\hat{r}\\+2\left((-p+\cos\gamma)\sin2\theta\right)\hat\theta
\end{align}

\noindent and

\begin{align}
\bs H_d^{\text{out}}(\bs r)=M_s\frac{R^4}{48r^4}(p-\cos\gamma)
\left[
    \left(\frac{1+3\cos 2\theta}{2}\right)\hat r
    +\sin2\theta\,\hat\theta
\right]\,.
\label{Hmagout}
\end{align}

One can notice that the value of $\gamma$ plays different roles in the magnetic field inside and outside the sphere. That is, in the region $r>R$, $\gamma$ contributes to a global factor to the magnetostatic field, which vanishes when the BP is spherically symmetric ($\gamma=0$ or $\pi$ for $p=1$ or $p=-1$, respectively). In such a case, the 
dipolar term in a multipolar expansion is expected to be null. As a consequence, hedgehog-like BPs can be only stabilized in the absence of dipolar energy, since the latter is proportional to $\propto \bs{m}\cdot \bs{H}_d$. On the other hand, because $\gamma$ does not contribute with a global factor in the magnetostatic field inside the sphere, even spherically symmetric BPs generate a dipolar field inside the nanoparticle. Indeed, a hedgehog BP pointing radially outside ($p=1$ and $\gamma=0$) generates a constant magnetostatic field pointing radially inside the sphere. This demagnetizing field yields the magnetic moments to deviate from the radial direction, and adopts a quasi-tangential configuration in the nanosphere equator \cite{Elias-EPL,Pyly-PRB}.

\begin{figure}
\centering
\includegraphics[width=9cm]{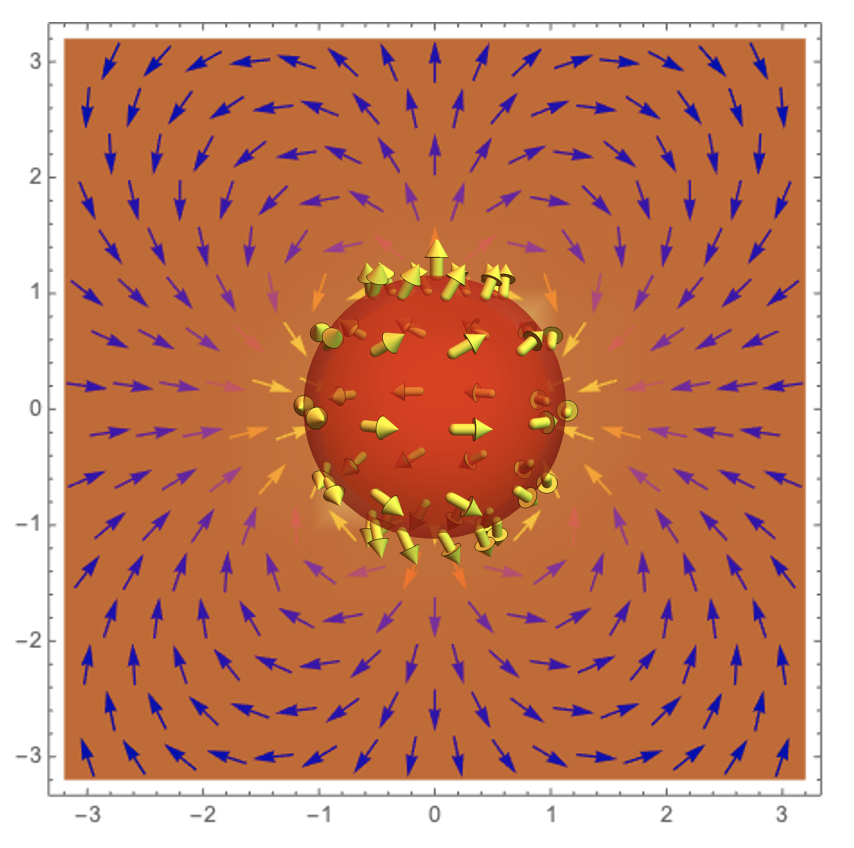}
\caption{(Normalized) External quadrupolar field produced by a BP hosted in a nanosphere. Here we have used $p = +1$ and $\gamma = \pi/2$.}
\label{H_field}
\end{figure}

The vector field of the magnetostatic field outside the sphere is depicted in Fig. \ref{H_field}, for $p=1$ and $\gamma=\pi/2$. From the analysis of the presented vector field and Eq. \eqref{Hmagout}, one notices the emergence of a magnetic field with quadrupolar nature, whose strength decreases with $\gamma$. This result is very interesting since quadrupole magnetic fields can be used, for instance, for trapping atoms in radio-frequency dressing experimental settings \cite{Mozitot}, designing magnetic tweezers to exert forces on
magnetic particles \cite{Zhang-IEEE}, and changing the properties of ferrofluid flow and heat transfer \cite{Alvarez}. The quadrupolar fields are generally obtained by two ferromagnetic bars parallel to each other, with the north pole of one next to the south of the other. The same field profile can be generated by two properly spaced coils with currents in opposite directions. Another interesting configuration of magnetic elements to obtain a quadrupolar field is considering four pole tips, with two opposing magnetic north poles and two opposing magnetic south poles. At the micro and nanoscale size, this last configuration can be observed in square spin ice systems \cite{Teonis,Mol,Loreto}, and magnetic tweezers \cite{Zhang-IEEE}. Therefore, the magnetic system considered in our work overcomes the necessity of using two or more magnetic elements to generate a quadrupole magnetic field. 

\end{section}

\section{Interaction between two magnetic spheres hosting Bloch point}

\begin{figure}
\centering
\includegraphics[width=8.5cm]{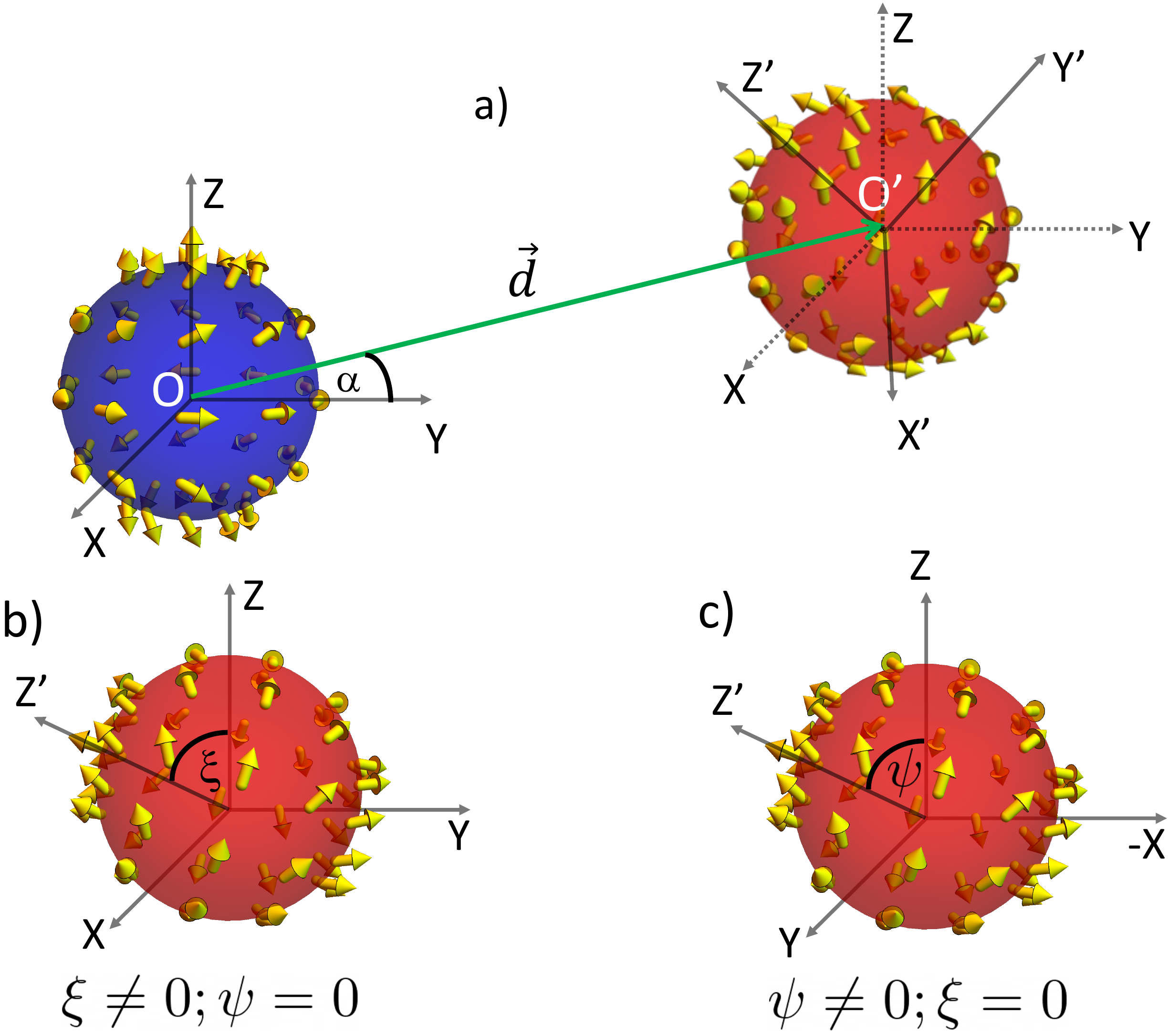}
\caption{Schematic representation of the idealized interaction between two nanospheres with a BP as a metastable state. In a), we assume that the blue BP (sphere 1), localized in the frame $O$, generates a magnetostatic field that allows the interaction with the red BP (sphere 2), localized in the frame $O'$. The vector $\vec{d}$ represents the separation between $O$ and $O'$, forming an angle $\alpha$ with the $Y$-axis. The BP in $O'$ is allowed to perform rigid rotations in response to the interaction energy, which are represented in b) and c) for two limiting cases.}
\label{fig_interaction}
\end{figure}

Once we have established the particular shape of the magnetostatic field generated by a Bloch Point confined in a nanosphere, we can explore how two BP's interact through the magnetostatic potential. Let us consider the interaction between two BP's hosted in spherical nanoparticles with the same radii $R_1=R_2=R$, separated by a distance $d$. We also consider that the magnetic spheres are not subject to any other force than that produced by the interaction between the magnetization field of one of them and the magnetic field produced by the other. Under these assumptions, the total magnetostatic energy $E_{m}$ of the system is given by

\begin{equation}\label{IntPrinc}
E_{m}=\frac{\mu_0}{2}\int \bs M\cdot \bs H,
\end{equation}

\noindent
where $\bs M(\bs r)=\bs M_1(\bs r)+\bs M_2(\bs r)$ and $\bs H$ is the magnetic field produced by the magnetic spheres. The situation can become very complex if the magnetic field is strong enough to modify the structure of the BPs. In this case, finding a new equilibrium BPs configuration would be necessary to minimize the system's energy. Due to this difficulty and the fact that the BP structure is mainly dominated by its self-dipolar field, we assume that the BPs structure do not change. This assumption can be justified by considering hard ferromagnetic materials in such a way that weak external fields cannot modify their configuration. In this approximation, Eq. \eqref{IntPrinc} can be simplified to

\begin{equation}
E_{\text{int12}}=\frac{\mu_0}{2}\int_{V_1}\bs M_1\cdot \bs H_2 \, dV_1+\frac{\mu_0}{2}\int_{V_2} \bs M_2\cdot \bs H_1\, dV_2\,,
\label{int_tot}
\end{equation}

\noindent
where the subindices 1 and 2 relates with the two BPs nucleated, respectively, in the blue and red spheres represented in Fig. \ref{fig_interaction}. 

From the reciprocity theorem \cite{Aharoni-Book}, we can state that both terms in the right-hand of the previous equation are equal. Therefore, we can rewrite Eq. \eqref{int_tot} as

\begin{equation}\label{ENINT}
E_{\text{int}}=\mu_0\int_{V_2}\bs M_2\cdot \bs H_1  dV_2.
\end{equation}

\noindent
In above equation, we consider that the BP stabilized in the sphere denoted by $1$ generates a magnetostatic field $\bs{H}_1$ in the region where the BP$_2$, which has magnetization field $\bs M_2$, is located in. Using $\nabla \cdot (U_{\text{d}} \bs M)=\bs M\cdot \nabla U_{\text{d}}+ U_{\text{d}}\nabla\cdot \bs M$, Eq. \eqref{ENINT} reads

\begin{align}
&E_{\text{int}} =\mu_0\int_{V_2}(U_{\text{d1}}\nabla\cdot \bs M_2- \nabla \cdot (U_{\text{d1}} \bs M_2)) dV_2\\
&=\mu_0\int_{V_2}U_{\text{d1}}\left(\nabla\cdot \bs M_2\right) dV_2- \mu_0\int_{S_2}\left(U_{\text{d1}} \bs M_2\right)\cdot \hat{\bs n}_2dS_2.
\label{eq_en_int}
\end{align}

It is worth noticing that Eq. \eqref{eq_en_int} is integrated out in the frame where the BP$_2$ is located. Therefore, we need to express the potential $U_{\text{d1}}$, generated by the BP$_1$, from the frame of the sphere $2$ (see Fig. \ref{fig_interaction}).  We have then determined a relationship between the two reference systems by assuming that, in the most general scenario, the two reference frames, $O$ and $O'$, are located in the center of the spheres $1$ and $2$, respectively. Additionally, we consider that the BPs axis coincides with the $Z$ and $Z'$-axis in each case. The reference frames are separated by a vector $\bs d$, pointing from $O$ to $O'$, as depicted in Fig. \ref{fig_interaction}. Also, sphere 2 ($O'$) is free to perform rigid rotations in response to interacting with the other BP. In this context, after some mathematical manipulations and considerations (see Appendix \ref{SupMat} for further details), we obtain the potential $U_{\text{d1}}$ in the frame $O'$, given by

\begin{align}
&\nonumber U_{\text{d1}} = \frac{c_1}{|\bs r+ \bs d|^3}\Bigg[\frac{3}{|\bs r+ \bs d\vert^2}\Big(d_z+r_2 \cos \theta_2 \cos \xi\cos\psi\\
&+r_2 \sin \theta_2 (\sin \xi  \cos \psi  \sin \phi_2 -\sin \psi  \cos \phi_2 )\Big)^2-1\Bigg],\hspace{0.7cm}
\end{align}

\noindent
where $c_1 = M_{s1}(p_1-\cos\gamma_1)R_1^4/12$ and we introduced the rotation angles $\xi$ and $\psi$, corresponding to two sucessive rotations around $X$ and $Y$, respectively.

A general result for the integrals presented in Eq. (\ref{eq_en_int}) becomes quite complicated. However, we can obtain interesting results by exploring some particular cases of interaction. Thus, we consider that the BP$_1$ (which generates de magnetostatic field) is characterized by $\gamma_1 = \arccos\left(-p_1/4\right)$ (we recall the fact that $\gamma = 0$ produces a null interaction energy) and fix its polarity $p_1$ pointing in the $Z-$axis direction. Similarly, for the BP$_2$, we adopt $\gamma_2=\arccos\left(-p_2/4\right)$ and consider that $p_2=\pm 1$. It is worth noting that because we assume that the interaction between the BPs does not change their helicities significantly, we use the $\gamma$ value for a BP free of external interactions \cite{Pyly-PRB}. Finally, we fix the position of BP$_1$ and vary the position of BP$_2$ along the plane $yz$. In this way, we explore three representative cases for the interaction between the BPs as a function of $\bs{d}=(0,d\cos\alpha,d\sin\alpha)$, where $\alpha$ is the angle between $\bs d$ and the axis-$Y$ (see Fig. \ref{fig_interaction}). Finally, we restrict our analysis to cases where the BPs interaction is appreciable, \textit{i. e.}, the separation between them is $\vert\bs{d}\vert= R_1+R_2+\epsilon$, being $\epsilon$ a small and positive parameter. The main idea is to describe how the magnetization field of the BP$_2$ changes in the presence of the magnetostatic potential $U_{\text{d1}}$. Therefore, we explore possible rigid rotations of the BP structure, which can be interpreted as variations of the direction to which the BP$_2$ magnetic moments in the sphere poles ($Z'$-axis) point. Thus, from determining the interaction energy as a function of the rotation angles $\psi$ and $\xi$, we obtain the respective rotation angle minimizes the energy or, equivalently, the direction to which $Z'$-axis points in order to minimize the system's energy. From now on, we numerically solve Eq. \eqref{eq_en_int} to determine the system's energy. In the obtained results, we adopt arbitrary unities, that is, $R=1$, $\mu_0=1$, $M_s=1$, $\epsilon=0.1$. Therefore, all the presented numerical results give us the qualitative behavior of the analyzed system.    

\subsection{Case 1: $\bs{d}=(0,d,0)$}

\begin{figure}
\centering
\includegraphics[width=8.5cm]{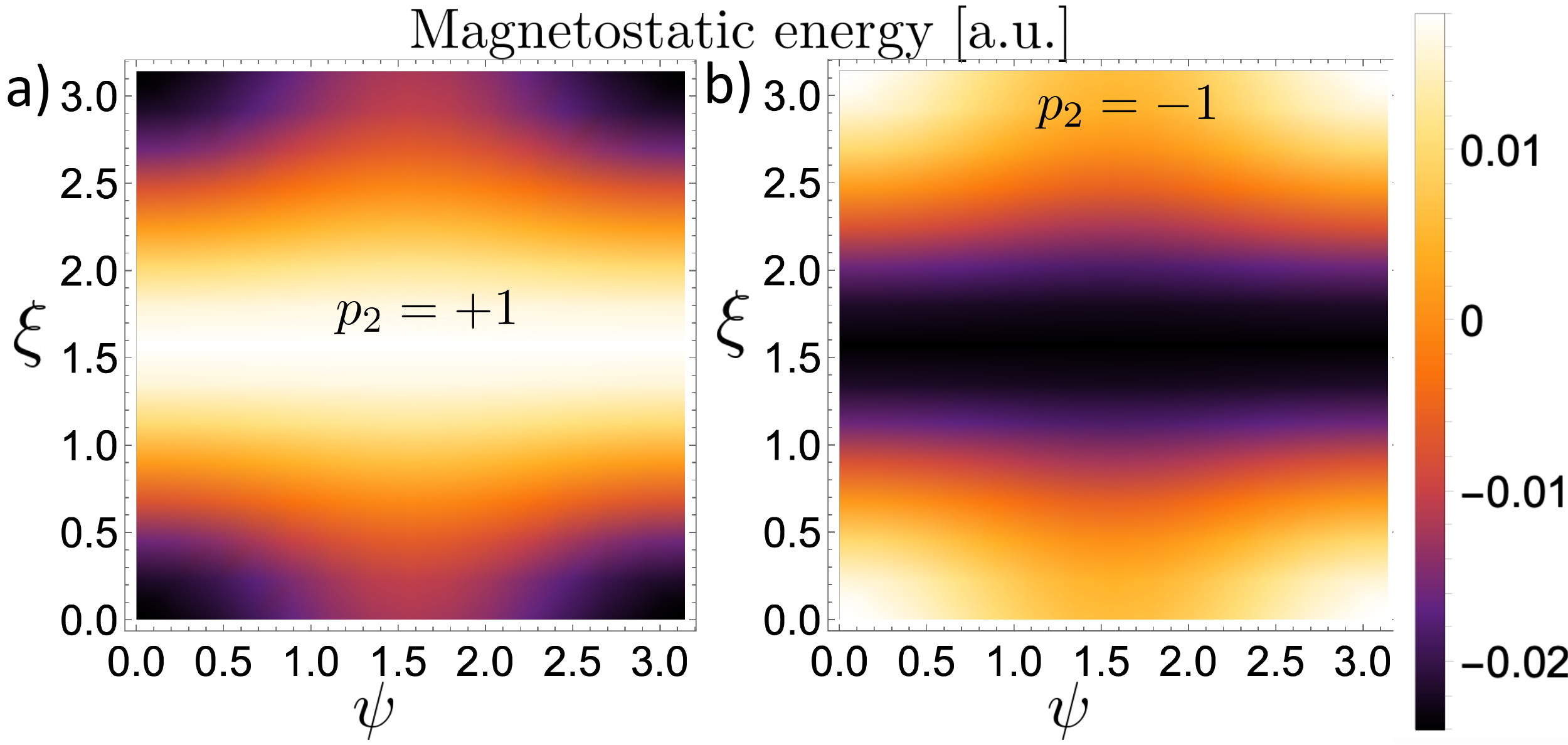}
\caption{Magnetostatic energy between two BPs interacting along the $Y$-axis, defined by $\alpha=0^{\circ}$. }
\label{fig:case1}
\end{figure}

\begin{figure}
\centering
\includegraphics[width=8.5cm]{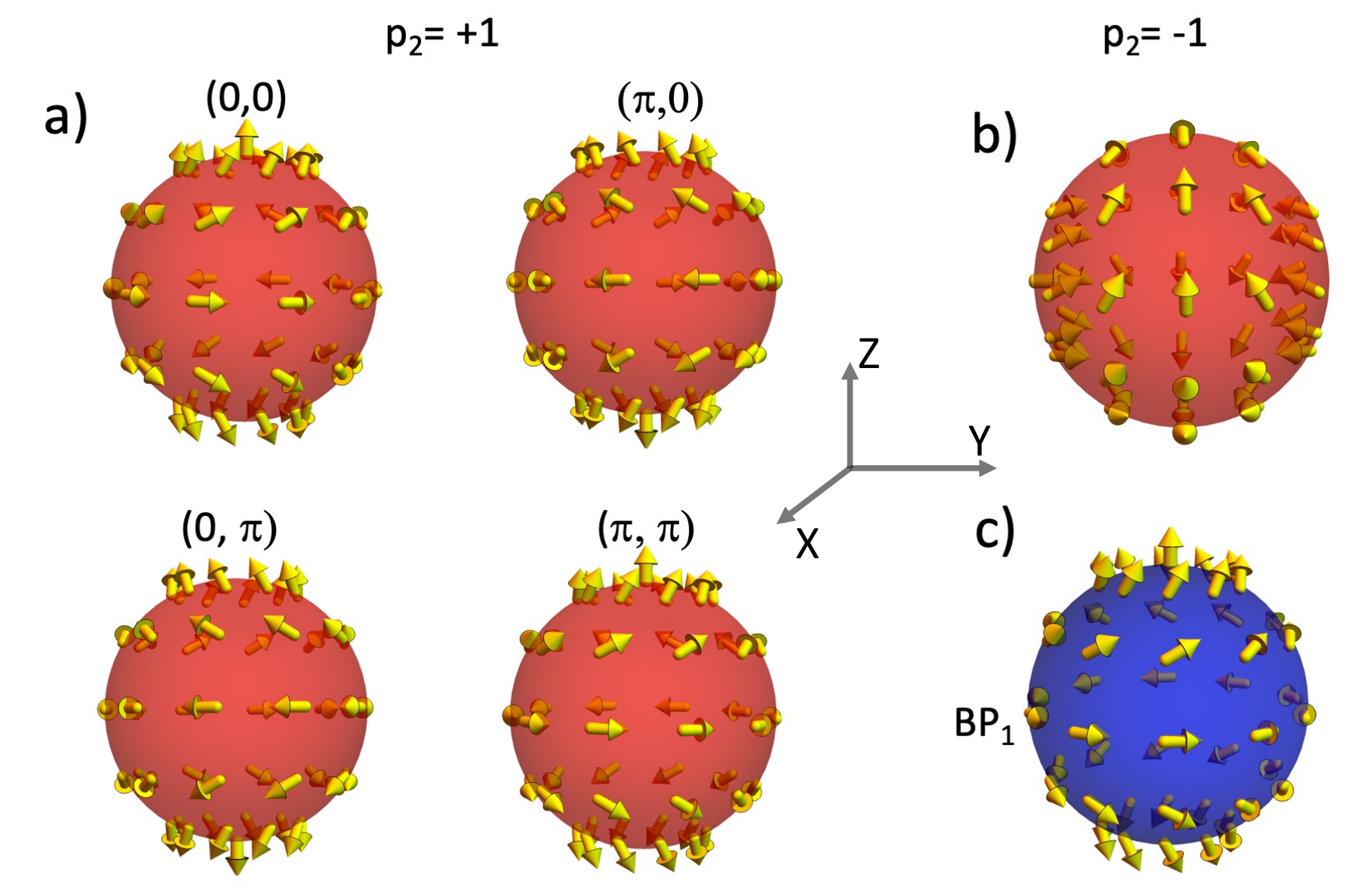}
\caption{The final state of the BP$_2$ magnetization configuration for the different rotation angles that minimize the magnetostatic energy in the case $\alpha=0^{\circ}$. Figures (a) and (b) correspond respectively to the polarities $p_2=1$ and $p_2=-1$. For comparison, the BP$_1$ profile is included in Figure (c), in blue color.}
\label{fig:FScase1}
\end{figure}

Firstly, we consider the interaction between two nanospheres hosting BPs lying in the plane $YZ$, along the $Y$-axis direction, that is, $\alpha = 0^{\circ}$. The obtained results are depicted in Fig. \ref{fig:case1}, where we present the magnetostatic energy (in arbitrary unities) as a function of the rotation angles $\psi$ and $\xi$ that determine the final direction to which the $Z'$-axis points. Fig. \ref{fig:case1}-(a) and (b) depict to the magnetostatic energy when the BP$_2$ has a polarity $p_2=+1$ and $p_2= -1$, respectively. An opposite behavior can be observed from the analysis of the obtained results. From Fig.\ref{fig:case1}-(a) we can see that the minimum energy is degenerated in four pairs of rotation angles $\{\psi,\xi\}$, namely {$\left\{(0,0),(\pi,0),(0,\pi),(\pi,\pi)\right\}$}. In this case, we can state that the BPs tend to align the magnetic moments in the sphere poles in the same direction ($\hat{Z}'=\pm\hat{Z}$). Additionally, the minimum energy is independent of the helicity orientation (clockwise or anticlockwise). Indeed, Fig. \ref{fig:FScase1}-(a) presents the magnetization field from each pair $\{\psi,\xi\}$ related to the minimum energy configuration. Its analysis reveals that, for $p_2 = +1$, the four degenerated states are reduced to only two since a rotation in $\xi=\pi$ essentially changes the sign of $\gamma$. Finally, when the BPs axis are oriented perpendicular to each other {($\xi=\pi/2$)}, we obtain the maximum energy values, which establishes an energy barrier between the distinct magnetic ground states for the considered system. In contrast, when $p_2=-1$, we obtain an opposite behavior. That is, the system minimizes its energy for $\xi=\pi/2$, whatever the rotation angle $\psi$, and the maximum values for the magnetostatic energy degenerate in the four pairs of rotation angles which minimize the energy in the case $p=+1$. The magnetization profile of the BP$_2$ in the minimum energy configuration is depicted in Fig. \ref{fig:FScase1}-(b). In Fig. \ref{fig:FScase1}-(c), we present the magnetization profile of BP$_1$ just for comparison with the final magnetic states of the BP$_2$.

\subsection{Case 2: $\bs{d}=(0,d,d)$}

\begin{figure}
\centering
\includegraphics[width=8.5cm]{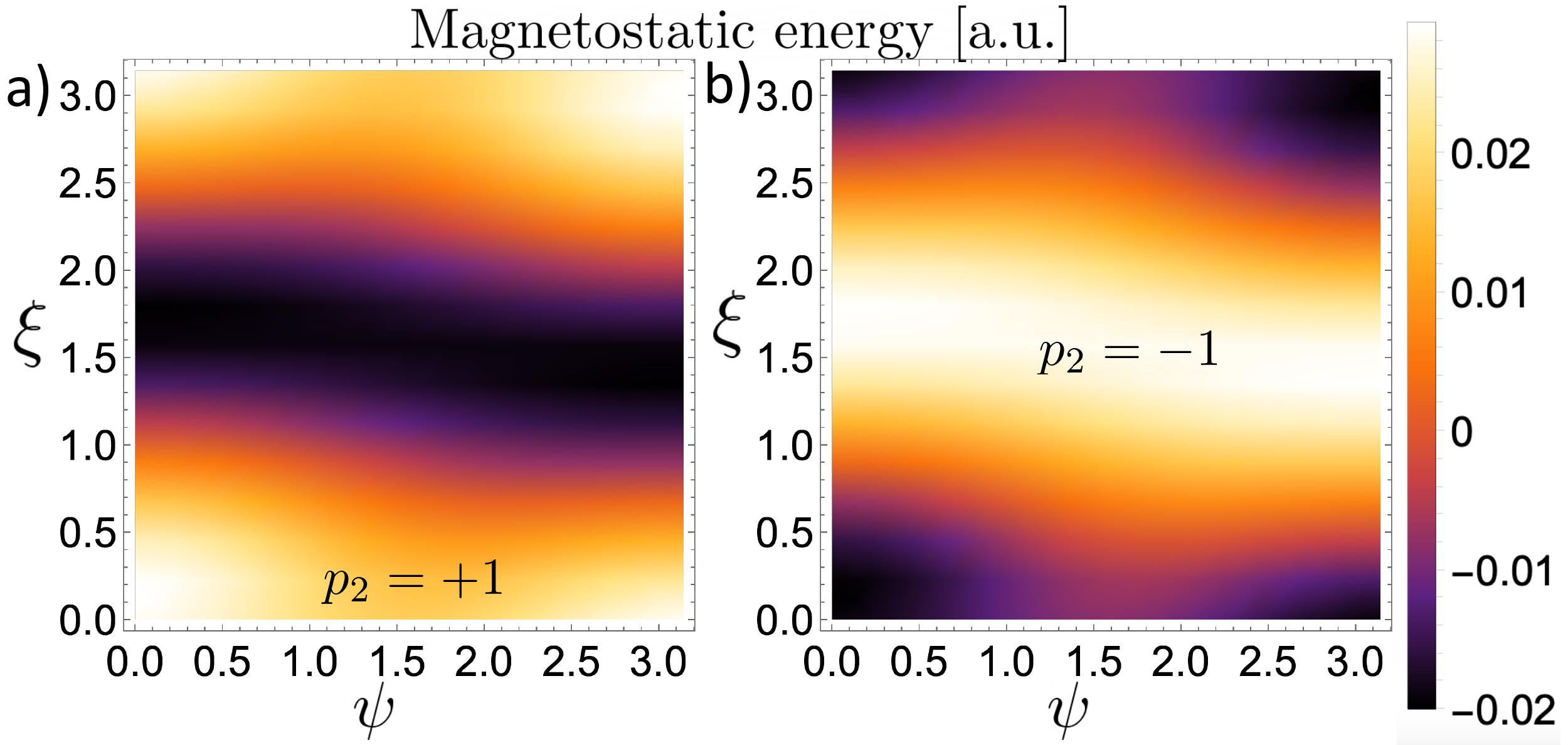}
\caption{Magnetostatic energy of a system composed by two BPs interacting along an axis defined by $\alpha=45^{\circ}$ as a function of the rotation angles $\psi$ and $\xi$.}
\label{fig:case2}
\end{figure}

\begin{figure}
\centering
\includegraphics[width=8.5cm]{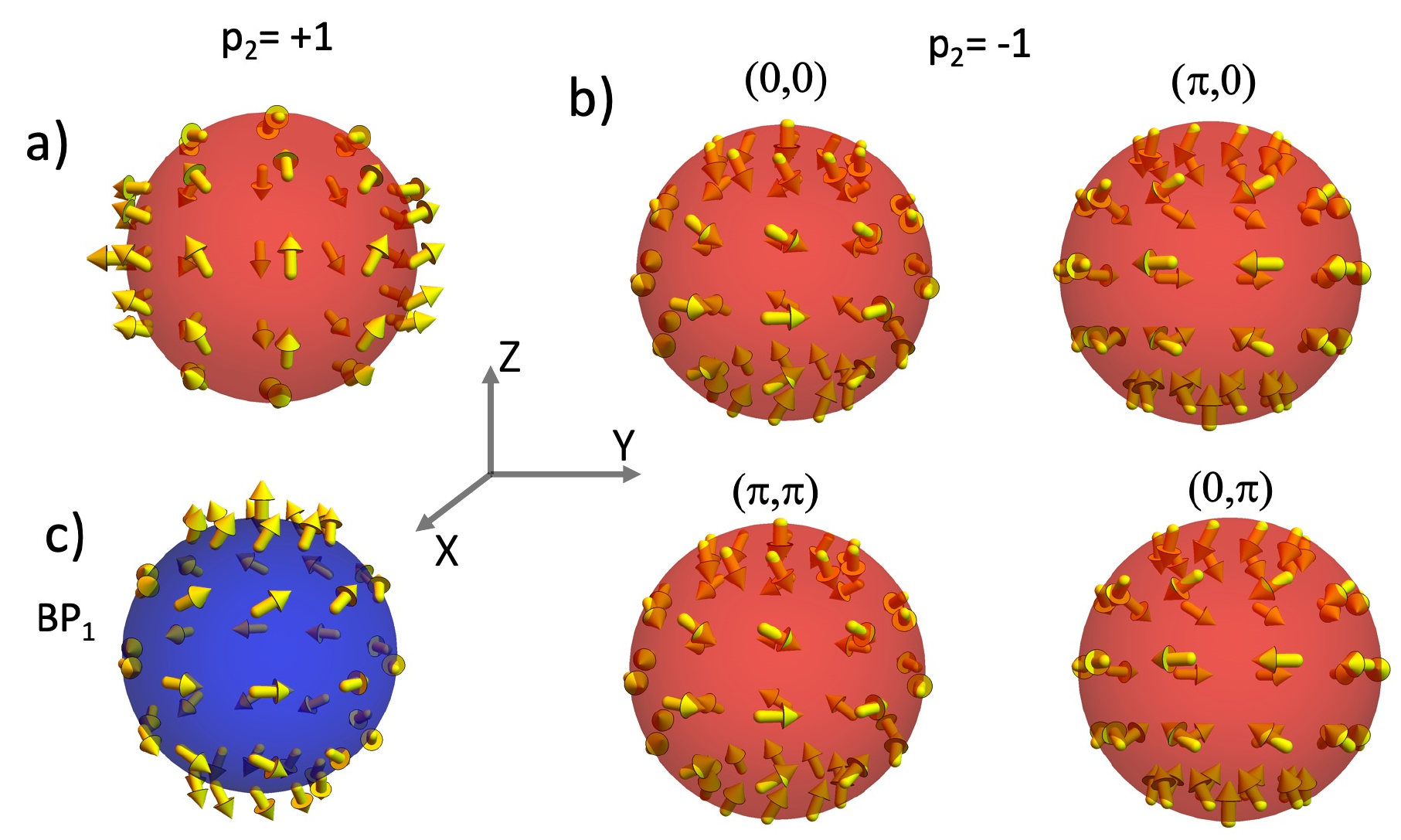}
\caption{Final configuration of the BP$_2$ for the different rotation angles that minimize the interaction energy in the case $\alpha=45^{\circ}$. Panels (a) and (b) correspond respectively to the cases where $p_2=1$ and $p_2=-1$. For comparison, the configuration of BP$_1$ was included in (c).}
\label{fig:FScase2}
\end{figure}

We continue our analysis on the magnetostatic interaction between two BPs hosted in nanospheres by considering that they lie in the $YZ$-plane, and their relative position forms an angle $\alpha = 45^{\circ}$. Our results evidence the same qualitative behavior as compared with the case of $\alpha=0^{\circ}$. That is, there are certain rotation angles where the minimum energy is reached, depending on the BP polarity and the specific orientations between the axes $Z$ and $Z'$. Nevertheless, due to the rotational symmetry of the BPs, the configuration that minimizes the energy when $\alpha=45^{\circ}$ is infinitely degenerated, as evidenced from the results presented in Fig. \ref{fig:case2}, which shows the magnetostatic energy as a function of the rotation angles $\psi$ and $\xi$. For each rotation $\psi$, around $X$-axis, there is an associated rotation $\xi$, around $Y$-axis, that leads the BPs to assume the minimum energy configuration. Indeed, the analysis of Fig. \ref{fig:case2}-(a) reveals that for $p_2=1$, there are infinite pairs of angles $\{\psi,\xi\}$ that minimize the energy. For instance, the minimum energy configuration of the BP$_2$ for the case $\psi=0$ is shown in Fig. \ref{fig:FScase2}-(a). Indeed, only one rotation around the $Y$-axis {($\xi=\pi/2$)} is allowed since any other value of $\xi$ would yield higher energy. Additionally, two pairs of rotation angles produce the configuration with maximum energy, e.g., {$(\psi,\xi)\sim (0,0)$ and $(\psi,\xi)\sim (\pi,\pi)$}, which are separated by the energy valley formed by the infinitely degenerated minimum energy states. If the BP$_2$ polarity is $p_2=-1$, one observes a similar behavior as the case where $p_2=1$ and $\alpha=0$. Indeed, because the inversion of the polarity leads to the opposite behavior in the magnetostatic interaction between the BPs, two pairs of rotation angles minimize the energy, as evidenced by the results presented in Fig. \ref{fig:case2}-(b). In this case, the pairs {$(\psi,\xi)\sim (0,0)$ and $(\psi,\xi)\sim (\pi,\pi)$} represent the configuration with minimum energy, which are separated by the energy barrier for the infinite number of pairs $(\psi,\xi)$. Nevertheless, it is worth noticing that although the points {$(\psi,\xi)\sim (\pi,0)$ and $(\psi,\xi)\sim (0,\pi)$} do not correspond to the minimum energy configuration, they can also be considered metastable states due to the pronounced energy barrier at the lines $\xi\sim\pi/2$. The possible configurations of BP$_2$ having minimum energy or metastable states are presented in Fig. \ref{fig:FScase2}-(b).

\subsection{Case 3: $\bs{d}=(0,0,d)$}

\begin{figure}
\centering
\includegraphics[width=8.5cm]{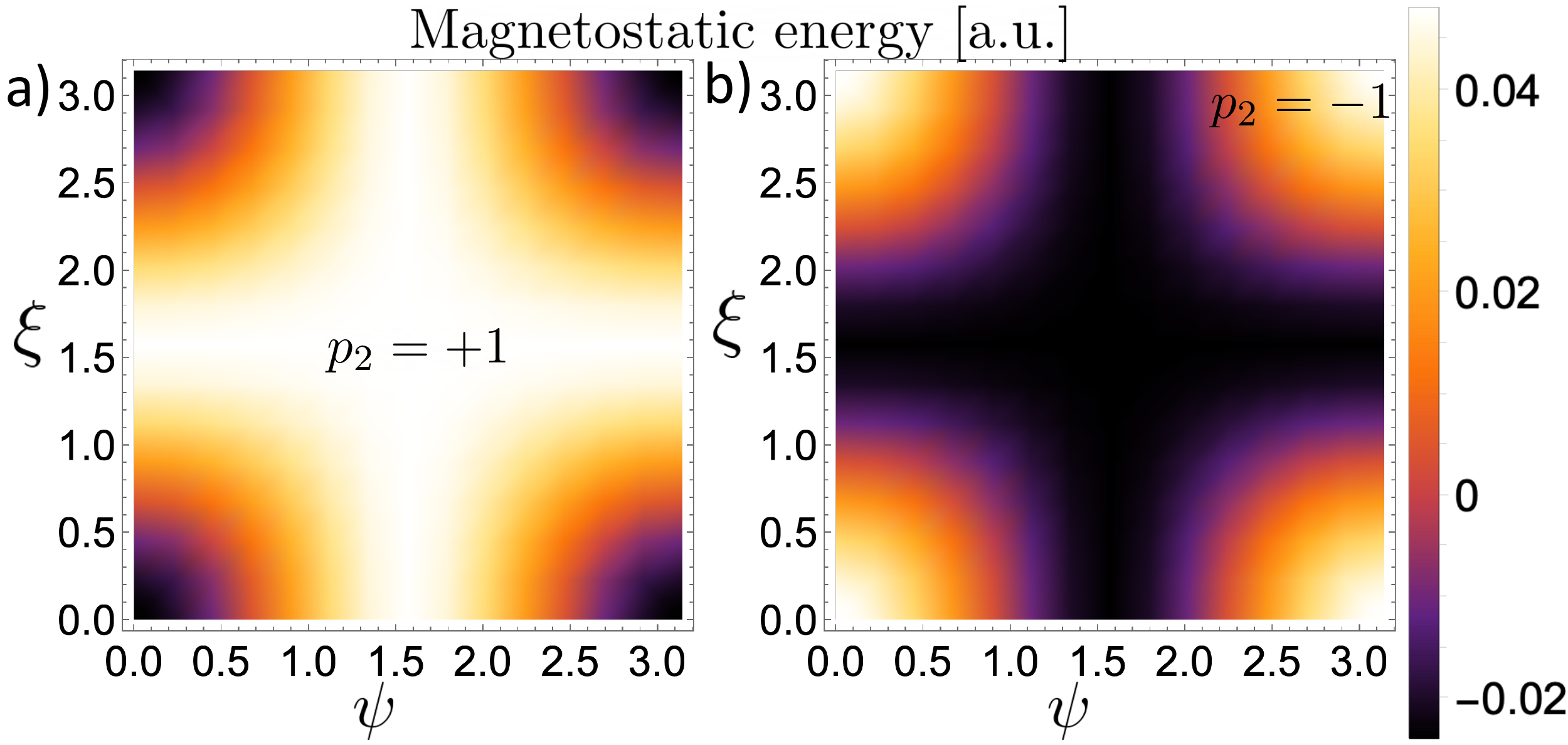}
\caption{Magnetostatic energy of a system composed by two BPs interacting along an axis defined by $\alpha=90^{\circ}$ as a function of the rotation angles $\psi$ and $\xi$}
\label{fig:case3}
\end{figure}

\begin{figure}
\centering
\includegraphics[width=8.5cm]{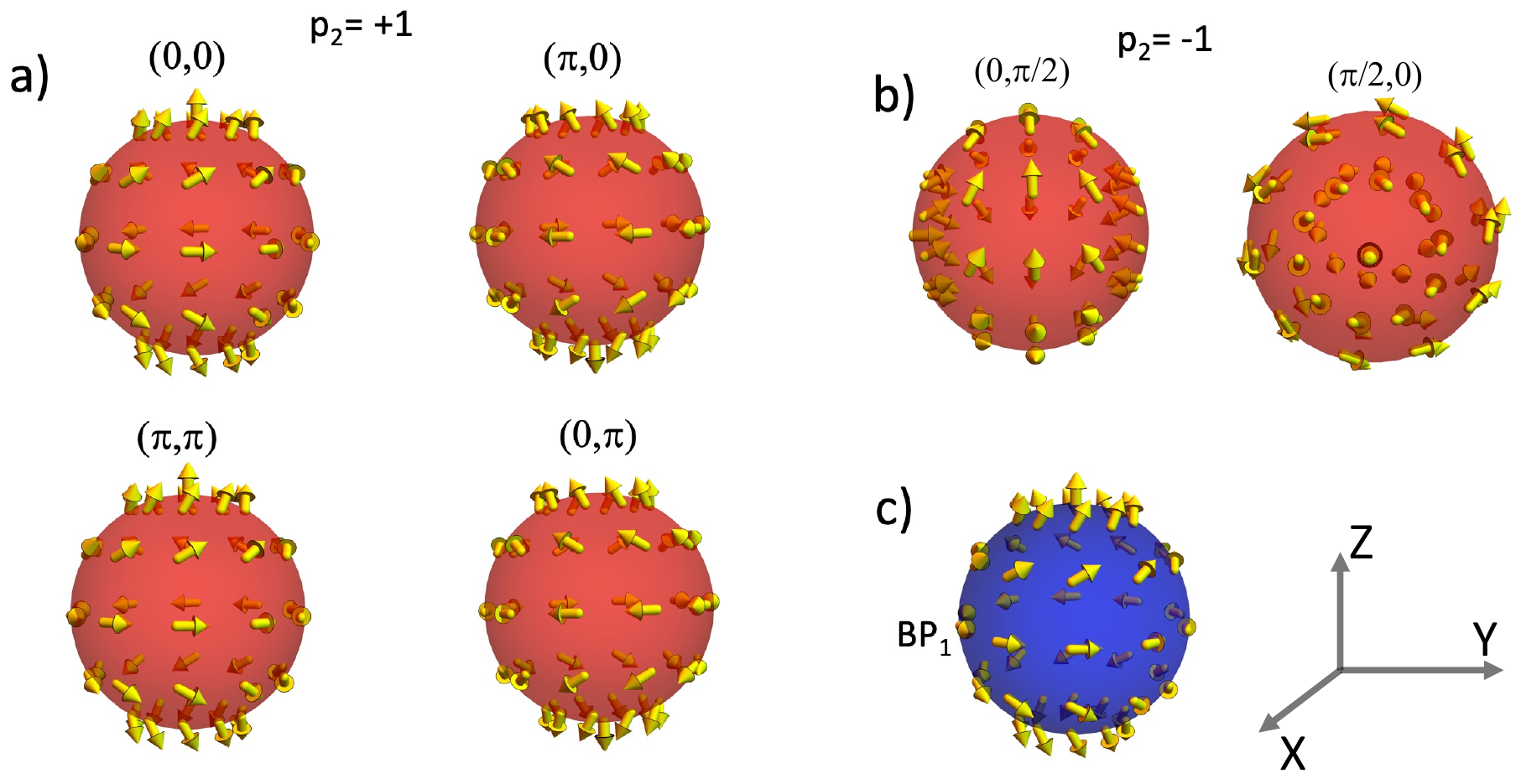}
\caption{Final configuration of the BP$_2$ for the different rotation angles that minimize the interaction energy in the case $\alpha=90^{\circ}$. Panels (a) and (b) correspond respectively to the cases where $p_2=1$ and $p_2=-1$. For comparison, the configuration of BP$_1$ was included in (c).}
\label{fig:FScase3}
\end{figure}

Finally, we consider the interaction between two BPs lying in the plane $YZ$, with their relative position forming the angle $\alpha = 90^{\circ}$. Under this assumption, the BP$_2$ is located at the $Z$-axis, just above the BP$_1$. We analyze the behavior of the magnetostatic energy as a function of possible BP$_2$ rotations around the axis $X$ and $Y$. The obtained results are presented in Fig \ref{fig:case3}-(a), from which one can notice that, for $p=+1$, the configuration having minimum energy is precisely the same as the case $\alpha=0^{\circ}$. That is, the minimum energy is degenerated in four pairs of rotation angles $\{\psi,\xi\}$, which can be reduced to only two rotations because the minimum energy is independent of the helicity orientation, as represented in Fig. \ref{fig:FScase3}-(a) in $\xi=\pi$ essentially changes the sign of $\gamma$. These minimum energy states are separated by an energy barrier whose maximum values are associated to an infinite number of possible rotations giver by {$(\psi,\pi/2)$ and $(\pi/2,\xi)$}, as depicted in Fig. \ref{fig:case3}-(b). Therefore, if {$\psi=\pi/2$}, the energy is independent of the rotation angle $\xi$. The same behavior is valid if we fix {$\xi=\pi/2$} and change the values of $\psi$. Two examples of the BP$_2$ configuration {($(0,\pi/2)$ and $(\pi/2,0)$)} producing the minimuum energy state are presented in Fig. \ref{fig:FScase3}-(b).

\subsection{Magnetostatic energy as a function of the distance}

\begin{figure}
\centering
\includegraphics[width=8.5cm]{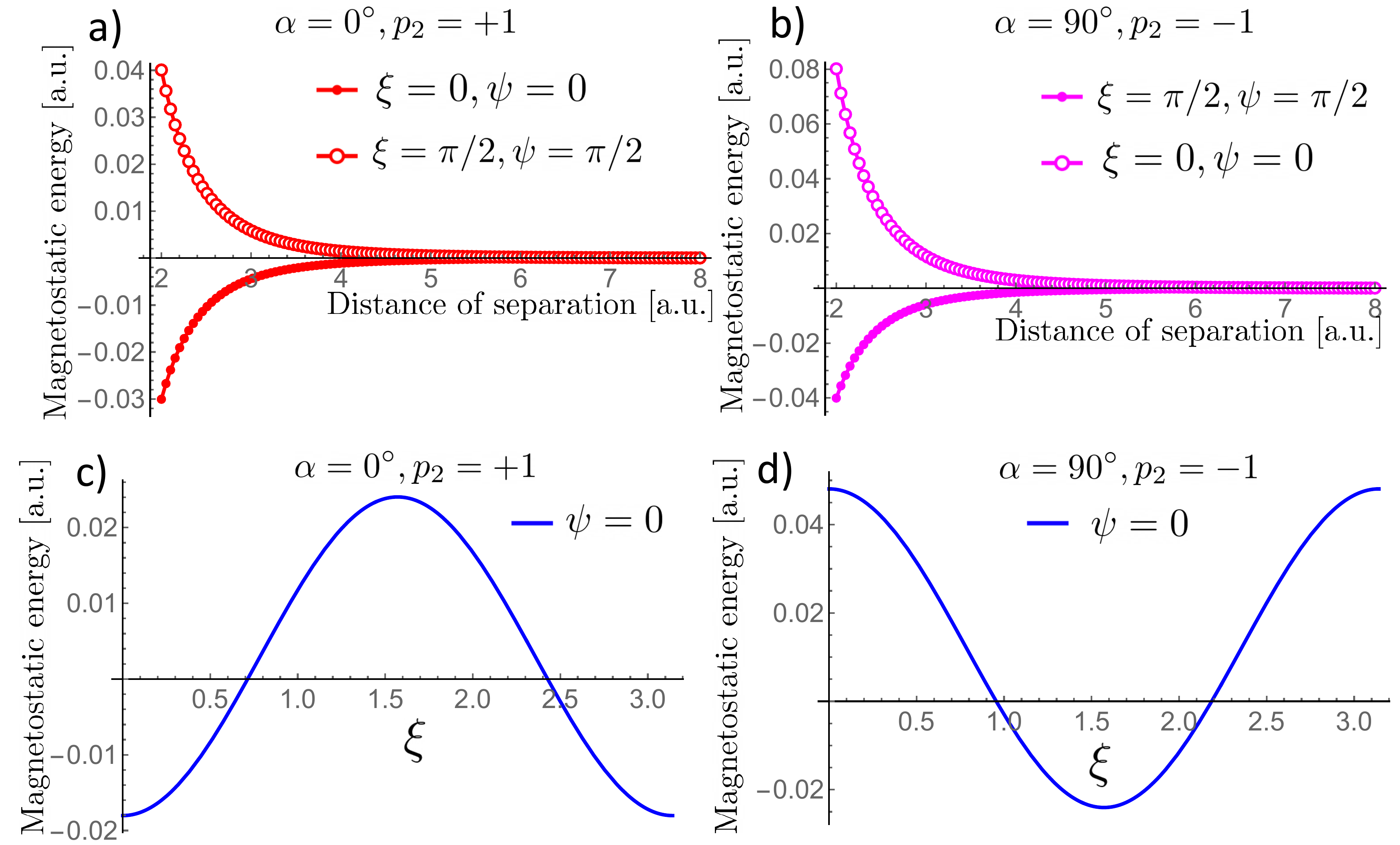}
\caption{Magnetostatic energy as a function of the BP distance. Panel (a) presents the results for $\alpha=0^{\circ}$ and $p_2=+1$, while panel (b) depicts the case $\alpha=90^{\circ}$ and $p_2=-1$. In both cases, open symbols correspond to the energy when the BPs are oriented to reach the maximum energy state, while filled symbols correspond to the energy when the BPs are oriented to reach the minimum energy configuration. Panel (c) depicts the behavior of the energy as a function of $\xi$ for a fix $\psi=0$ in the case $\alpha=0^{\circ}$ with $p_2=+1$. Panel (d) shows the corresponding energy as a function of $\xi$ for $\psi=0$ in the case $\alpha=90^{\circ}$ and $p_2=-1$.}
\label{Evsd}
\end{figure}

The above-analyzed cases present the same qualitative behavior regarding the magnetostatic interaction as a function of the rotations of BP$_2$ concerning BP$_1$. From this perspective, we study the nature of these interactions if they are attractive or repulsive. We have then determined the magnetostatic energy of the system composed of the two BPs as a function of the distance between them. To clarify the role of the relative orientation between the BPs on the interaction, we present the results for the magnetostatic energy when the BPs are oriented to reach the minimum and maximum energy state for the cases $\alpha=0^{\circ}$ with $p_2=+1$ (Fig. \ref{Evsd}-(a)), and $\alpha=90^{\circ}$ with $p_2=-1$ (Fig. \ref{Evsd}-(b)). Although we present only these two particular cases, the same qualitative behavior is observed in all analyzed cases under the respective combinations of $\alpha$ and $p_2=\pm 1$, yielding maximum and minimum energy configurations. Our results show that the magnetic configurations presenting minimum and maximum energies have attractive and repulsive natures. Indeed, the analysis of Figs. \ref{Evsd}-(a) and \ref{Evsd}-(b) reveals that the energy increases with the nanospheres distance when the relative orientation of the BPs leads to the minimum energy configuration, evidencing an attractive interaction. On the other hand, a decrease in the magnetostatic energy is observed when the BPs are oriented at an angle defining the maximum energy configuration, revealing, in this case, the repulsive nature of the BPs interaction.  

Because the relative rotation between the BPs changes the nature of the interaction from attractive to repulsive, there should be a set of specific rotation angles for which the BPs do not interact. To corroborate this statement, we calculate the magnetostatic energy numerically as a function of $\xi$ for a fixed $\psi$ and distance between the nanospheres. The obtained results are depicted in Figs. \ref{Evsd}-(c) and \ref{Evsd}-(d) respectively for $\alpha=0^{\circ}$ and $\alpha=90^{\circ}$, where we fix $\psi=0^{\circ}$. In specific, for the case $\alpha=0^{\circ}$ with $p_2 = +1$ the system becomes noninteracting when $\xi \sim 41^{\circ} $ or $\xi \sim 138^{\circ} $. Similarly, in the case $\alpha= 90^{\circ}$ with $p_2=-1$, the BPs do not interact when $\xi\sim 54^{\circ} $ or $\xi\sim 125^{\circ}$. As expected, the qualitative behavior for both cases is similar, but the specific angles for which the interaction vanishes depend on $\alpha$.

\begin{section}{Discussion and conclusions}

The knowledge of the magnetostatic field generated by magnetic nanoparticles is an important issue to consider when analyzing the interaction between magnetic particles. In this work, we analyzed the magnetic field of nanospheres hosting a BP as a metastable state. Our results evidenced the quadrupolar nature of the magnetostatic field generated by a BP. Furthermore, the strength of this field depends on the BP helicity, vanishing for a hedgehog BP pattern. This result is interesting from both perspectives, fundamental and applied physics. Indeed, the magnetic system considered in our work overcomes the necessity of using two or more magnetic elements to generate a quadrupole magnetic field, which can give a new breath for technological propositions regarding quadrupole magnetic fields \cite{Mozitot,Zhang-IEEE,Alvarez}.

The magnetostatic interaction between two nanospheres with a BP metastable state is also determined. Our results evidenced that the interaction between BPs depends on the relative position of one BP regarding the other. Additionally, the relative orientation between the BP axis determines the strength of the magnetostatic energy. Depending on the BPs polarities, there are relative orientations of the BPs magnetic moments for which the energy is minimized. Nevertheless, the energy minima are separated by relative rotation angles for which the energy is a maximum. In this context, although the interaction between planar vortices (2D counterparts of BPs) can be analyzed from the perspective of their topological charges \cite{Belo}, we can conclude that the interaction between BPs should consider arguments beyond the approximation of topological charge-mediated interaction. Indeed, BPs with the same topological charge ($p_1=p_2=+1$) can interact with attractive or repulsive potential, depending on the relative rotation angle between them. The same behavior is valid for BPs with opposite topological charges ($p_1=-p_2$). Finally, because the nature of the interaction can change from attractive to repulsive, we showed that there are relative orientations between the BPs for which they do not interact. These results evidence that nanospheres with BPs as magnetic states should present an exciting interaction dynamic, highly depending on the initial conditions. Although the results presented in this work concern the magnetostatic interaction of BPs, they give some insights in possible ways to understand the complex behavior of three-dimensional topological textures.

\end{section}
\begin{section}{Acknowledgments}
N. V-S acknowledges funding from Fondecyt Iniciacion No. 11220046. ASN acknowledges funding from Fondecyt Regular 1190324, and Financiamiento Basal  para  Centros  Cient\'ificos  y  Tecnol\'ogicos  de  Excelencia AFB180001. Powered@NLHPC: This research
was partially supported by the supercomputing infrastructure of the NLHPC (ECM-02). V.L.C.-S. thanks the financial support of the Brazilian agencies CNPq (Grant n. 302084/2019-3) and Fapemig (Grant n.APQ-0648-22). The work of F. Tejo was supported by ANID + Fondecyt de Postdoctorado, convocatoria 2022 + Folio 3220527.
\end{section}

\clearpage
\onecolumngrid
\appendix

\section{Appendix}\label{SupMat}

\subsection{Magnetostatic field generated by the BP in the whole space}

In this Supplementary Material, we explicitly show the calculation of the magnetostatic field generated by the BP in the whole space. We recall the definition of the demagnetizing field, given by $\bs{H}_d(\bs r) = -\nabla U_d(\bs{r})$, where

\begin{align}
U_d(\bs{r})=\frac{M_s}{4\pi}\left( - \int_{V'} \frac{\nabla\cdot \bs m(\bs r')}{|\bs r-\bs r'|}dV'+ \int_{S'} \frac{\bs m(\bs r')\cdot \bs n'}{|\bs r-\bs r'|}dS'\right).
\label{sm:potential}
\end{align}

Additionally, the normalized magnetization field, expressed in spherical coordinates, is ${\bs m}=\left(\sin\Theta({\bs r})\cos\Phi({\bs r}),\sin\Theta({\bs r})\sin\Phi({\bs r}),\cos\Theta({\bs r})\right)$. To describe the BP magnetic profile, we use the following ansatz already presented in the main text

\begin{align}
    \label{sm:ansatz}
    \Theta(\theta) = p\theta + \pi(1-p)/2\\
    \Phi(\phi) = q\phi+\gamma,\,.
\end{align}

To perform our calculations, it is convenient to write the BP magnetic profile explicitly in terms of the components of the magnetization field in the spherical basis, that is

\begin{align}
    m_r = \cos\gamma\sin^2\theta + p\cos^2\theta\\
    m_{\theta}=\sin\theta\cos\theta\left(\cos\gamma -p\right)\\
    m_{\phi} = \sin\gamma\sin\theta.
    \label{sm:components}
\end{align}

\noindent
Thus, the volumetric and superficial magnetic charges are evaluated as

\begin{align}
    \nabla\cdot\bs{m} = \frac{1}{r}\left[\left(\cos^2\theta + 1\right)\cos\gamma+p\sin^2\theta\right]\\
    \bs{m}\cdot\bs{n}=\cos\gamma\sin^2\theta+p\cos^2\theta.
    \label{sm:charges}
\end{align}

Additionally, due to the spherical symmetry of the nanoparticle geometry, we expand the Green's function $G(\bs{r},\bs{r}') = \vert\bs{r}-\bs{r}'\vert^{-1}$ into the spherical harmonics basis

\begin{align}
    \frac{1}{\vert\bs{r}-\bs{r}'\vert}=\sum_{l=1}^{\infty}\sum_{m=-l}^l\frac{4\pi}{2l+1}\frac{r_<^l}{r_>^{l+1}}Y_{l,m}(\theta,\phi)Y_{l,-m}(\theta',\phi'),
    \label{sm:green}
\end{align}

\noindent
where $Y_{l,m}(\theta,\phi)$ is a spherical harmonic function of degree $l$ and order $m$, formally defined as

\begin{align}
    Y_{l,m}=\sqrt{\frac{2l+1}{4\pi}\frac{(l-m)!}{(l+m)!}}P_l^m(\cos\theta)e^{im\phi},
    \label{sm:spheharm}
\end{align}

\noindent
being $P_l^m(x)$ the associated Legendre polynomials, and $r_{<}(r_>)$ corresponds to the smaller (larger) value between $r$ and $r'$. By substituting Eqs. (\ref{sm:ansatz} - \ref{sm:spheharm}) into Eq. (\ref{sm:potential}), and after some algebraic manipulation, we obtain that the magnetostatic potential is given by

 \begin{align}
 \label{sm:potential2}
     U_d(\bs r) = \nonumber\frac{M_s}{2}\sum_{l=1}^{\infty}P_l(\cos\theta)\Bigg[\frac{r_<^l}{r_>^{l+1}}R^2\left(\frac{1}{3}(2\cos\gamma+p)\frac{2\delta_{0,l}}{2l+1}+\frac{2}{3}(p-\cos\gamma)\frac{2\delta_{2,l}}{2l+1}\right)\\
     -\int_0^R\frac{r_<^l}{r_>^{l+1}}r'dr'\left(\frac{2}{3}(2\cos\gamma+p)\frac{2\delta_{0,l}}{2l+1}-\frac{2}{3}(p-\cos\gamma)\frac{2\delta_{2,l}}{2l+1}\right) \Bigg],
 \end{align}
 
 \noindent
where $\delta_{\alpha,\beta}$ is the Kronecker delta. In the following, we solve Eq. (\ref{sm:potential2}) for the regions inside and outside the sphere. 

\subsubsection{Region $r \leq R$ (inside the sphere)}
Here we have to properly evaluate $r_<$ and $r_>$ in the region of interest. For the first term of Eq. (\ref{sm:potential2}) is evident that $r_< = r$ and $r_>=R$. However, in the second term the radial integral involved in Eq. (\ref{sm:potential2}) must be carefully evaluated. Indeed, it can be written in a more convenient way to explicitly recognize $r_<$ and $r_>$

\begin{align}
    \int_0^R\frac{r_<^l}{r_>^{l+1}}r'dr'=\frac{1}{r^{l+1}}\int_0^rr'^{l+1}dr'+r^l\int_r^R\frac{dr'}{r'^l}=\frac{2l+1}{(l+2)(l-1)}r-\frac{1}{l-1}\frac{r^l}{R^{l-1}}.
\end{align}

\noindent
 Thus, by substituting above equation in Eq. ($\ref{sm:potential2}$) and performing some algebra, we obtain the potential inside the sphere, given by
 
 \begin{align}
    U_d^{\text{in}}(\bs r) = \frac{M_s}{24}\left(9pr-8pR+15r\cos\gamma - 16R\cos\gamma + 3r\cos2\theta(p-\cos\gamma)\right)
 \end{align}
 
 \noindent
 which corresponds to Eq. (\ref{Udin}) in the main text. Note that the above result coincides with Ref. \cite{andreas}. Next, the corresponding magnetostatic field reads
 
 \begin{align}
\bs{H}_d^{\text{in}}(\bs r)=     -\nabla U_d^{\text{in}}(\bs r)=-M_s\frac{R^4}{8}\Big[\left(3p+5\cos\gamma+(p-\cos\gamma)\cos2\theta\right)\hat{r}+2\left((-p+\cos\gamma)\sin2\theta\right)\hat\theta\Big],
 \end{align}
 
 \noindent
 which is Eq. (\ref{Hmagin}) in the main text.

  \subsubsection{Region $r  > R$ (outside the sphere)}
  In this case, it is always satisfied that $r_<=r'$ and $r_>=r$. Therefore, the integral involved in the second term of Eq. (\ref{sm:potential2}) is
  
  \begin{align}
    \int_0^R\frac{r_<^l}{r_>^{l+1}}r'dr'=\frac{1}{r^{l+1}}\int_0^Rr'^{l+1}dr'=\frac{1}{l+2}\frac{R^{l+2}}{r^{l+1}}.
  \end{align}

From the substitution of above equation in Eq. (\ref{sm:potential2}), we obtain 

\begin{align}
    U_d^{\text{out}}(\bs r)=\frac{M_s(p-\cos\gamma)}{12}\left(3\cos^2\theta-1\right)\frac{R^4}{r^3},
    \label{Udoutappen}
\end{align}

\noindent
whose corresponding magnetostatic field is given by 

\begin{align}
    \bs H_d^{\text{out}}(\bs r)=M_s\frac{R^4}{48r^4}(p-\cos\gamma)
\left[
    \left(\frac{1+3\cos 2\theta}{2}\right)\hat r
    +\sin2\theta\,\hat\theta
\right],
\end{align}

\noindent
which correspond to Eq. (\ref{Hmagout}) in the main text.

\subsection{Magnetostatic potential represented in the frame $O'$}
As stated in the main text, the way we chose to explore the magnetostatic interaction between two spheres hosting a BP as a metastable state is by locating a BP$_1$ in a frame $O$, which generates a magnetostatic field $\bs{H}_{d1}^{\text{out}}=-\nabla U_{\text{d1}}^{\text{out}}$. Then a BP$_2$, situated in a frame $O'$, interacts with the BP$_1$ through the integral presented in Eq. \ref{eq_en_int} of the main text. Therefore, the problem is reduced to express $U_{\text{d1}}^{\text{out}}$ in terms of the $O'$-variables, i.e., as seen from the frame $O'$. To calculate the magnetostatic potential in the frame $O'$, we can transform coordinates from one system to the other as

\begin{equation}
\bs r - \bs d = \mathcal{R}\bs r',
\end{equation}

\noindent
where $\mathcal{R}$ is the rotation from the system $O$ to $O'$, and $\bs d = (d_x,d_y,d_z)$ is a vector pointing from $O$ to $O'$, whose magnitude is the distance between the center of the spheres. Because BPs are axial, we need only two rotation angles to relate the frames with each other. Let's consider the rotation $\mathcal{R}=Y_\psi X_\xi$, rotation of angle $\xi$ around $X$ and then a rotation of $\psi$ around $Y$. That is

\begin{equation}
\mathcal{R}=\left(
\begin{array}{ccc}
    \cos\psi & \sin\xi\sin\psi&\cos\xi\sin\psi\\
    0 & \cos\xi&-\sin\xi\\
    -\sin\psi & \cos\psi\sin\xi&\cos\xi\cos\psi
\end{array}
\right).
\end{equation}

From the above considerations, we have that the magnetostatic potential generated by the BP$_1$ at the origin of $O$ is

\begin{align}
U_{\text{d1}}&=\frac{c_1}{r_1^3}(3 \cos^2\theta_1-1),
\end{align}

\noindent
where $c_1 = \frac{M_{s1}}{12}(p_1-\cos\gamma_1)R_1^4$ (see Eq. \ref{Udoutappen}) and the sub-indices $1,2$ are used to referring to frames $O$ and $O'$, respectively. Now, using the relation

\begin{equation}
    \bs r_1=\bs d + \mathcal{R}\bs r_2,
\end{equation}

\noindent
we can compute the interaction energy at the frame $O'$. Indeed, the relation of the polar angles between both frames reads

\begin{equation}
    \cos\theta_1=\frac{z_1}{r_1}=\frac{d_z+r_2 \cos \theta_2 \cos \xi  \cos \psi +r_2 \sin \theta_2 (\sin \xi  \cos \psi  \sin \phi_2 -\sin \psi  \cos \phi_3 )}{|\mathcal{R}\bs r_2+ \bs d|}\,,
\end{equation}

\noindent
where

\begin{eqnarray}
    \mathcal{R}\bs r_2 =\left(
\begin{array}{ccc}
    d_x+r_2\cos\phi_2\cos\psi\sin\theta_2 + r_2\sin\psi\left(\cos\theta_2\cos\xi+\sin\theta_2\sin\xi\sin\phi_2\right) \\
    d_y-r_2\cos\theta_2\sin\xi+r_2\cos\xi\sin\theta_2\sin\phi_2 \\
    d_z+r_2\cos\theta_2\cos\xi\cos\psi+r_2\sin\theta_2\left(\cos\psi\sin\xi\sin\phi_2 - \cos\phi_2\sin\psi\right)
\end{array}
\right). 
\end{eqnarray}

Therefore, the magnetostatic potential generated by the BP$_1$ seen from the system 2 ($O'$) becomes

\begin{equation}
U_{\text{d1}}= \frac{c_1}{|\mathcal{R}\bs r_2+ \bs d|^3}\left[3\left(\frac{d_z+r_2 \cos \theta_2 \cos \xi  \cos \psi +r_2 \sin \theta_2 (\sin \xi  \cos \psi  \sin \phi_2 -\sin \psi  \cos \phi_2 )}{|\mathcal{R}\bs r_2+ \bs d|}\right)^2-1\right]\,,
\end{equation}

\noindent
from which we can formally calculate the expression for $E_\text{int}$ in Eq. (\ref{eq_en_int}). 

\begin{thebibliography}{0}%
\makeatletter
\providecommand \@ifxundefined [1]{%
 \@ifx{#1\undefined}
}%
\providecommand \@ifnum [1]{%
 \ifnum #1\expandafter \@firstoftwo
 \else \expandafter \@secondoftwo
 \fi
}%
\providecommand \@ifx [1]{%
 \ifx #1\expandafter \@firstoftwo
 \else \expandafter \@secondoftwo
 \fi
}%
\providecommand \natexlab [1]{#1}%
\providecommand \enquote  [1]{``#1''}%
\providecommand \bibnamefont  [1]{#1}%
\providecommand \bibfnamefont [1]{#1}%
\providecommand \citenamefont [1]{#1}%
\providecommand \href@noop [0]{\@secondoftwo}%
\providecommand \href [0]{\begingroup \@sanitize@url \@href}%
\providecommand \@href[1]{\@@startlink{#1}\@@href}%
\providecommand \@@href[1]{\endgroup#1\@@endlink}%
\providecommand \@sanitize@url [0]{\catcode `\\12\catcode `\$12\catcode
  `\&12\catcode `\#12\catcode `\^12\catcode `\_12\catcode `\%12\relax}%
\providecommand \@@startlink[1]{}%
\providecommand \@@endlink[0]{}%
\providecommand \url  [0]{\begingroup\@sanitize@url \@url }%
\providecommand \@url [1]{\endgroup\@href {#1}{\urlprefix }}%
\providecommand \urlprefix  [0]{URL }%
\providecommand \Eprint [0]{\href }%
\providecommand \doibase [0]{http://dx.doi.org/}%
\providecommand \selectlanguage [0]{\@gobble}%
\providecommand \bibinfo  [0]{\@secondoftwo}%
\providecommand \bibfield  [0]{\@secondoftwo}%
\providecommand \translation [1]{[#1]}%
\providecommand \BibitemOpen [0]{}%
\providecommand \bibitemStop [0]{}%
\providecommand \bibitemNoStop [0]{.\EOS\space}%
\providecommand \EOS [0]{\spacefactor3000\relax}%
\providecommand \BibitemShut  [1]{\csname bibitem#1\endcsname}%
\let\auto@bib@innerbib\@empty
\end{thebibliography}%


\begin{thebibliography}{99}


\bibitem{Parkin-Nat}
S. Parkin and S.-H. Yang, Nat. Nano. \textbf{10}, 195 (2015).

\bibitem{Fert-Nat}
A. Fert, N. Reyren, and V. Cros, Nature \textbf{2}, 17031 (2017).

\bibitem{Hrcak}
G. Hrkac, J. Dean, and D. A. Allwood, 
Philos. Trans. A Math. Phys. Eng. Sci. \textbf{369}, 3214 (2011).

\bibitem{Grolier}
J. Grollier, D. Querlioz, K. Y. Camsari, K. Everschor-Sitte, S. Fukami, and M. D. Stiles, 
Nat. Electron. \textbf{3}, 360 (2020).

\bibitem{Marrows-APL}
C. H. Marrows and K. Zeissler, 
Appl. Phys. Lett. \textbf{119}, 250502 (2021).

\bibitem{Rajaraman}
R. Rajaraman: \textit{Solitons and Instantons} (North-Holland, Amsterdam, 1984).

\bibitem{Sampaio-Nat}
J. Sampaio, V. Cros, S. Rohart, A. Thiaville, and A. Fert, 
Nat. Nanotechnol. \textbf{8}, 839 (2013).

\bibitem{Skyrmionium}
X. Zhang, J. Xia, Y. Zhou, D. Wand, X. Liu, W. Zhao, and M. Ezawa, 
Phys. Rev. B \textbf{94}, 094420 (2016).

\bibitem{biskyrmion}
X. Yu, Y. Tokunaga, Y. Kaneko, Y. Matsui, Y. Tagushi, and Y. Tokura, 
Nat. Commun. \textbf{5}, 3198 (2014).

\bibitem{bimerons-1}
B. G\"obel, A. Mook, J. Henk, I. Mertig, and O. A. Tretiakov, 
Phys. Rev. B \textbf{99}, 060407 (2019).

\bibitem{bimerons-2}
A. S. Ara\'ujo, R. J. C. Lopes , V. L. Carvalho-Santos, A. R. Pereira, R. L. Silva, R. C. Silva, and D. Altbir, 
Phys. Rev. B \textbf{102}, 104409 (2020).

\bibitem{bimerons-3}
N. Gao, S.-G. Je, M.-Y. Im, J.W. Choi, M. Yang, Q. Li, T.Y. Wang, S. Lee, H.-S. Han, K.-S. Lee, W. Chao, C. Hwang , J. Li, and Z.Q. Qiu, 
Nat. Commun. \textbf{10}, 5603 (2019).

\bibitem{bimerons-4}
M.V. Sapozhnikov, D.A. Tatarskiy, and V.L. Mironov, 
J. Magn. Mag. Mat. \textbf{549}, 169043 (2022).

\bibitem{Donnely-PRL}
C. Donnelly, M. Guizar-Sicairos, V. Scagnoli, M. Holler, T. Huthwelker, A. Menzel, I. Vartiainen, E. M\"uller, E. Kirk, S. Gliga, J. Raabe, and L. J. Heyderman, 
Phys. Rev. Lett. \textbf{114}, 115501 (2015).

\bibitem{Amalio-SciRep}
A. Fern\'andez-Pacheco, L. Serrano-Ram\'on, J. M. Michalik, M. R. Ibarra, J. M. De Teresa, L. O'Brien, D. Petit, J. Lee, and R. P. Cowburn, 
Sci. Rep. \textbf{3}, 1492 (2013).

\bibitem{May-NatCom}
A. May, M. Saccone, A. van den Berg, J. Askey, M. Hunt, and S. Ladak, 
Nat. Comm. \textbf{12}, 3217 (2021).

\bibitem{Phatak-PRL}
C. Phatak, A. K. Petford-Long, and M. De Graef, 
Phys. Rev. Lett. \textbf{104}, 253901 (2010).

\bibitem{Oksana}
E. Berganza, J. A. Fernandez‑Roldan, M. Jaafar, A. Asenjo, K. Guslienko, and O. Chubykalo‑Fesenko, 
Sci. Rep. \textbf{12}, 3426 (2012).

\bibitem{Birsh-ACS}
M. T. Birch, D. Cort\'es-Ortu\~no, L. A. Turnbull, \textit{et al.}, 
Nat. Commun. \textbf{11}, 1726 (2020).


\bibitem{Seki-NatMat}
S. Seki, M. Suzuki, M. Ishibashi, R. Takaji, N. D. Khanh, Y. Shiota, K. Shibata, W.Koshibae, Y. Tohura, and T. Ono, 
Nat. Mater. \textbf{21}, 181–187 (2022).


\bibitem{Donnely-NatPhys}
C. Donnelly, K. L. Metlov, V. Scagnoli, M. Guizar-Sicairos, M. Holler, N. S. Bingham, J. Raabe, L. J. Heyderman, N. R. Cooper, and Sebastian Gliga, 
Nat. Phys. \textbf{17}, 316 (2021).

\bibitem{bobbers-1}
F. N. Rybakov, A. B. Borisov, S. Bl\"ugel, and N. S. Kiselev, 
Phys. Rev. Lett. \textbf{115}, 117201 (2015).

\bibitem{bobbers-2}
F. Zheng, F. N. Rybakov, A. B. Borisov, D. Song,
S. Wang, Z.-A. Li, H. Du, N. S. Kiselev, J. Caron, Andr\'as Kov\'acs, M. Tian, Y. Zhang, S. Bl\"ugel, and R. E. Dunin-Borkowsk, 
Nat. Nanotech. \textbf{13}, 451 (2018).

\bibitem{Tai-PRL}
J.-S. B. Tai and I. I. Smalyukh, Phys. Rev. Let. \textbf{121}, 187201 (2018).

\bibitem{Liu-PRB}
Y. Liu, R. K. Lake, and J. Zang, 
Phys. Rev. B \textbf{98}, 174437 (2018).

\bibitem{Castillo-PRB}
S. Castillo-Sep\'ulveda, R. Cacilhas, V. L. Carvalho-Santos, R. M. Corona, and D. Altbir, 
Phys. Rev. B \textbf{104}, 184406 (2021).

\bibitem{Sutcliffe-PRL}
P. Sutcliffe, 
Phys. Rev. Lett. \textbf{118}, 247203 (2017). 

\bibitem{Col-PRB}
S. Da Col, S. Jamet, N. Rougemaille, A. Locatelli, T. O. Mentes, B. S. Burgos, R. Afid, M. Darques, L. Cagnon, J. C. Toussaint, and O. Fruchart, 
Phys. Rev. B \textbf{89}, 180405 (2014).

\bibitem{Im-NatCom}
M.-Y. Im, H.-S. Han, M.-S. Jung, Y.-S. Yu, S. Lee, S. Yoon, W. Chao, P. Fischer, J.-I. Hong, and K.-S. Lee, 
Nat. Commun. \textbf{10}, 593 (2019).

\bibitem{Feldkeler}
R. Feldtkeller, Z. Angew. Phys. \textbf{19}, 530 (1965).

\bibitem{Doring}
W. D\"oring, J. Appl. Phys. \textbf{39}, 1006 (1968).

\bibitem{Galkina}
E.G. Galkina, B.A. Ivanov, and V.A. Stephanovich, J. Magn. Mag. Mat. \textbf{118}, 373 (1993).

\bibitem{Thiaville-PRB}
A. Thiaville, J. M. Garcia, R. Dittrich, J. Miltat, and T. Schrefl,
Phys. Rev. B \textbf{67}, 094410 (2003).

\bibitem{Hertel-PB}
R. Hertel and J. Kirschner, Physica B: Condensed Matter 343,
206 (2004).

\bibitem{Hertel-PRL}
R. Hertel, S. Gliga, M. Fahnle, and C. M. Schneider, Phys. Rev.
Lett. \textbf{98}, 117201 (2007).

\bibitem{Milde-SC}
P. Milde, D. K\"ohler, J. Seidel, L. M. Eng, A. Bauer, A. Chacon,
J. Kindervater, S. M\"uhlbauer, C. Pfleiderer, S. Buhrandt,
C. Sch\"utte, and A. Rosch, Science \textbf{340}, 1076 (2013).

\bibitem{Beg-SciRep}
M. Beg, R. A. Pepper, D. Cort\'es-Ortu\~no, B. Atie, M.-A. Bisotti, G. Downing, T. Kluyver, O. Hovorka, and H. Fangohr, Sci. Rep. \textbf{9}, 7959 (2019).

\bibitem{Saez-RP}
G. S\'aez, P. D\'iaz, N. Vidal-Silva, J. Escrig, and E. E.Vogel, Results in Physics \textbf{39}, 105768 (2022)

\bibitem{Wieser-PRB}
R. Wieser, U. Nowak, and K. D. Usadel, Phys. Rev. B \textbf{69}, 064401 (2004).

\bibitem{Jamet-Book}
S. Jamet, N. Rougemaille, J.C. Toussaint, and O. Fruchart, ``\textit{25 - head-to-head domain walls in one-dimensional nanostructures: An extended phase diagram ranging from strips to cylindrical wires},'' in Magnetic Nano- and Microwires, Woodhead Publishing Series in Electronic and Optical Materials, edited by
Manuel V\'azquez (Woodhead Publishing, 2015) pp. 783 - 811.

\bibitem{Moreno-JMMM}
R. Moreno, V.L. Carvalho-Santos, D. Altbir, and O. Chubykalo-Fesenko, J. Magn. Mag. Mat. \textbf{542}, 168495 (2022).

\bibitem{Saez-2}
G. S\'aez, E. Saavedra, N. Vidal-Silva, J. Escrig, and Eugenio E.Vogel, Results in Physics \textbf{37}, 105530 (2022).

\bibitem{Elias-PRB}
R. G. El\'ias, V. L. Carvalho-Santos, A. S. N\'u\~n ez, and A. D. Verga, Phys. Rev. B \textbf{90}, 224414 (2014).

\bibitem{Carvalho-AOP}
V.L. Carvalho-Santos, R.G. El\'ias, and A.S. Nunez, Ann. Phys. \textbf{363}, 364 (2015).

\bibitem{Jia-Nat}
C. Jia, D. Ma, A. F. Sch\"affer, and J. Berakdar, Nat. Comm. \textbf{10}, 2077 (2019).

\bibitem{Ma-APL}
X.-P. Ma, J. Zheng, H.-G. Piao, D.-H. Kim, and P. Fischer, Appl. Phys. Lett. \textbf{117}, 062402 (2020).

\bibitem{Elias-EPL}
R.G. El\'ias, and A. Verga, Eur. Phys. J. B \textbf{82}, 159 (2011).

\bibitem{Pyly-PRB}
O. V. Pylypovskyi, D. D. Sheka, and Y. Gaididei, Phys. Rev. B \textbf{85}, 224401 (2012).

\bibitem{Tejo-SciRep}
F. Tejo, R. H. Heredero, O. Chubykalo‑Fesenko, and K. Y. Guslienko, Sci. Rep. \textbf{11}, 21714 (2021).

\bibitem{Hairy}
 P. Renteln, \textit{Manifolds, Tensors, and Forms: An Introduction for Mathematicians and Physicists}. Cambridge Univ. Press (2013).

\bibitem{Aharoni-Book}
A. Aharoni, \textit{Introduction to the Theory of Ferromagnetism}. Oxford Science Publications (1998).

\bibitem{Mozitot}
O. Morizot, C. L. Garrido Alzar, P.-E. Pottie, V. Lorent, and H. Perrin, J. Phys. B: At. Mol. Opt. Phys. \textbf{40}, 4013 (2007).

\bibitem{Zhang-IEEE}
Z. Zhang, K. Huang, and C.-H. Menq, IEEE/ASME Trans. on Mechatronics \textbf{15}, 704 (2010).

\bibitem{Alvarez}
S. A. Zonouzi, R. Khodabandeh, H. Safarzadeh, H. Aminfar, Y. Trushkina, M. Mohammadpourfard, M. Ghanbarpour, G. S. Alvarez, Experimental Thermal and Fluid Science
\textbf{91}, 155 (2018).

\bibitem{Teonis}
T. S. de Paiva, J. H. Rodrigues, L. A. S. M\'ol, A. R. Pereira, J. Borme, P. P. Freitas, and C. I. L. de Araujo, Sci. Rep. \textbf{10}, 9959 (2020).

\bibitem{Mol}
L. A. M\'ol, R. L. Silva, R. C. Silva, A. R. Pereira, W. A. Moura-Melo, and B. V. Costa, J. Appl. Phys. \textbf{106}, 063913 (2009).

\bibitem{Loreto}
R. S. Gon\c calves, R. P. Loreto, T. S. de Paiva, J. Borme, P. P. Freitas, and C. I. L. de Araujo,  Appl. Phys. Lett. \textbf{114}, 142401 (2019).


\bibitem{Belo}
L. R. A. Belo, N. M. Oliveira-Neto, W. A. Moura-Melo, A. R. Pereira, and E. Ercolessi, Phys. Lett. A \textbf{365}, 463 (2007).

\bibitem{andreas}
Andreas, Ch., Forschungszentrum Jülich, Vol. \textbf{88} (2014).

































%




















\end{thebibliography}
\end{document}